\documentclass[fleqn,10pt]{wlscirep}
\usepackage[utf8]{inputenc}
\usepackage[T1]{fontenc}
\title{Nucleation of cadherin clusters on cell-cell interfaces}

\author[*]{Neil Ibata and Eugene M. Terentjev}
\affil{Cavendish Laboratory, University of Cambridge, Cambridge, CB3 0HE, U.K.}
\affil[*]{emt1000@cam.ac.uk}

\begin{abstract}
Cadherins mediate cell-cell adhesion and help the cell determine its shape and function. Here we study collective cadherin organization and interactions within cell-cell contact areas, and find the cadherin density at which a `gas-liquid' phase transition occurs, when cadherin monomers begin to aggregate into dense clusters. We use  a 2D lattice model of a cell-cell contact area, and coarse-grain to the continuous number density of cadherin to map the model onto the Cahn-Hilliard coarsening theory. This predicts the density required for nucleation, the characteristic length scale of the process, and the number density of clusters. The analytical predictions of the model are in good agreement with experimental observations of cadherin clustering in epithelial tissues.
\end{abstract}
\begin{document}

\flushbottom
\maketitle
\thispagestyle{empty}


\section*{Introduction}

Many eukaryotic cells use membrane-bound integrin adhesion clusters \cite{Wozniak2004} to tether themselves to the extra-cellular matrix surrounding them (ECM). Similarly, these cells use membrane-bound cadherin adhesion clusters \cite{VanRoy2008,Niessen2007,Harris2010} to bind to their neighbouring cells directly. Adhesion molecules mediate mechanical signalling between the cell and its exterior by participating in important intracellular signalling pathways \cite{Burridge1996,Geiger2009,Yap2003,Cavallaro2004}. Clusters of adhesion molecules also help determine the structure of the cell; these shape changes are essential if the cell is to topologically fit into a tissue (e.g. in dividing epithelia \cite{LeBras2014}), to change its function (fibroblasts differentiating into myofibroblasts when placed on stiff media \cite{Goffin2006,Solon2007}), or to move (during wound healing \cite{Gates1994,Koivisto2014} or cancer metastasis \cite{Nagano2012,Eke2015}). In order to understand why any of these processes occur, we must first understand why there are clusters of adhesion molecules at all -- in physical terms, how their nucleation from a uniform distribution of sensors occurs.

Density-dependent nucleation is ubiquitous in soft matter. Changes in the concentration of attractively coupled molecules can help form large-scale symmetry-breaking structures. 
Computational studies have investigated the clustering of both integrin \cite{Gottschalk2004,Paszek2009} and cadherin \cite{Chen2016,Thompson2020}. 
The classical theory of aggregation on fluctuating membranes \cite{Weikl2002,Weikl2009,Speck2010} explains the unstable growth of nuclei into large-scale receptor domains (evident in low-resolution experiments), notably including the case of large cadherin domains on vesicles \cite{Fenz2017}. These effects are driven by the weak long-range forces mediated by membrane fluctuations, and cannot account for the stability of small nanometer-scale clusters of the kind recently identified in high-resolution cadherin imaging \cite{Wu2015,Indra2018}. We recently analytically investigated the nucleation of integrins in the high-concentration limit on the edge of a spreading cell \cite{Ibata2020}. Here, we extend this approach to the nucleation of such small punctate cadherin clusters on a generic cell-cell contact surface, stabilized by the strong short-range bonding between monomers.

We will first review background literature to examine the parallels between the aggregation of the types of adhesion molecules. Next, we build a lattice gas model to obtain the Ginzburg-Landau free energy for fluctuations in the density of cadherin molecules on the realistic 2D contact plane between cells. We consider whether our model correctly predicts the size and spacing of punctate adherens junctions seen in experiment.


 \textbf{Integrin/Cadherin analogy.} \ The cell needs to develop adhesion clusters in order to correctly remodel its shape in response to external substrates and mechanical cues \cite{Gumbiner1996,Parsons2010}. The large clusters visible under microscope (called focal adhesions for integrins \cite{Goffin2006,Gov2006}, and zonula adherens for cadherin \cite{Meng2009,Choi2016}) are the end-product of many smaller clusters aggregating over minutes or hours.

\begin{figure*}
 \centering
\includegraphics[width=\textwidth]{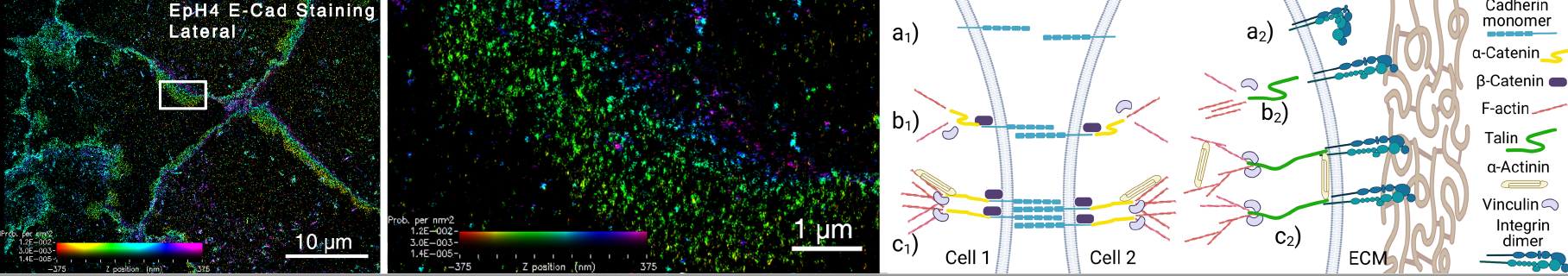}
\caption{Left panel: E-cadherin (`E' standing for epithelial) aggregate into punctate clusters with a characteristic size $\lesssim 50$ nm and separation $\gtrsim 100$ nm. Reproduced with permission from Wu et al. \cite{Wu2015}. Right panel: Aggregation of cadherins (1) and integrins (2). a) Single adhesion molecules sit in the cell membrane, inactive. b$_1$) Catenins begin to aggregate around the transmembrane cadherins, which form transmembrane \textit{trans}-bonds, and link to the cytoskeleton \cite{Niessen2007,Bajpai2008}: an individual adhesion complex forms. b$_2$) Integrins are activated by pulling when an individual adhesion complex assembles \cite{Liu2000}. c) Adhesion units aggregate laterally via cadherin \textit{cis}-bonds (1) or auxiliary proteins for integrin (2), and form clusters.}\label{fig:cadherin_sketch}
\end{figure*}

In order to be stabilized, both integrins \cite{Kong2009} and cadherins \cite{Rakshit2012} rely on `catch bonds' \cite{Thomas2008}, which strengthen under load. These bonds can help activate both molecules and are often preceded by the formation of a larger cytoplasmic protein complex which links an individual adhesion molecule to the cytoskeleton (see Fig.~\ref{fig:cadherin_sketch}). Integrin uses talin and vinculin to bind to F-actin \cite{Humphries2007,Ziegler2008}, while cadherin primarily relies on catenins \cite{Hartsock2008,Niessen2008}. After the protein complex is fully assembled, the acto-myosin cytoskeleton exerts pulling force on the adhesion molecule, strengthening the catch bond. Individual complexes can then aggregate into growing clusters, and link the cytoskeleton with the outside of the cell over a larger area, spreading the force applied by the actomyosin cortex \cite{Balaban2001,Shemesh2005}. Provided that adhesion molecules cluster together in sufficient numbers to withstand the load, the cytoskeleton can develop increasingly large pulling forces and distort the shape of the cell \cite{Burridge2016}.

Cadherins differ most notably from integrins in their ability to multimerise. {Cadherins can form two types of bonds with other cadherins: \textit{trans} bonds that link cadherins from two different cells \cite{Chitaev1998,Ozawa2002}, which effectively mimic the integrin binding to extra-cellular matrix, and \textit{cis} bonds that link neighbouring cadherins within the same cell, in the plane of the membrane.}
While integrins and cadherins might develop a very similar force-chain with the cytoskeleton in the direction normal to the cell membrane, their interactions within the plane of the cell membrane are fundamentally different. Whereas integrins appear to use secondary molecules to link to each other (possibly $\alpha$-actinin \cite{Otey1993,Sampath1998,Kelly2005,Roca-Cusachs2013}), cadherins bind directly to each other \cite{Wu2010}. 

The distance between cadherin monomers in a loosely packed cadherin lattice is ca.$7$ nm \cite{Harrison2011,Nagar1996,Lambert2005}, but can be as small as $3$ nm in very tightly packed junctions \cite{Thompson2019}; both of these figures are much smaller than the separation than between neighbouring integrins: ca.$30$ nm \cite{Xu2016}. The \textit{cis} bond strength can be estimated by extrapolating from measurements of cadherin \textit{trans} bond strength and the dissociation rates of these bonds, as well as numerical simulations of cadherin clustering. Note that it is much faster than the dissociation of cadherins in well-developed adhesion junctions with many such bonds established \cite{Li2013}.

Over the last ten years, the classical view has shifted from the idea that \textit{cis}-dimerisation preceded \textit{trans}-dimerisation \cite{Fichtner2014}. The discovery of an intermediate crossed-dimer, called the X-dimer, which precedes a more stable swapped \textit{trans}-dimerisation suggests that \textit{cis}-dimerisation is not necessary for cadherins to form stable cross-membrane bonds \cite{Wu2010,Harrison2011}. More recent work \cite{Pontani2016} suggests that cadherin cross-membrane dimers form before \textit{cis} clustering occurs. This can be further quantified by examining recent work on the development of large adherent regions in giant unilamellar vesicles (GUVs). In particular,  Fenz et al. \cite{Fenz2017}  showed that cadherin \textit{trans}-bonding, a characteristic of the extended adhesive region, spread from nucleation sites if membrane fluctuations were sufficiently small. In Supplementary Part A, we show that the size of membrane fluctuations within an adhesive region in a cell is much smaller than that required for adhesive regions to separate into separate domains in this model, based on the data from \cite{Biswas2017,DosSantos2016}. This means that within an adhesive region, a large portion of cadherins are indeed within \textit{trans}-bonded dimers (with \textit{on}-rates $\ll 1$ s, given an intrinsic lifetime of $0.63$ s \cite{Rakshit2012}). Lateral interactions between cadherin \textit{trans}-bonded pairs are small in this model (a few $k_B T$ at most), relying on membrane fluctuations to slowly and randomly bring cadherin pairs together. In contrast, small crystalline clusters of cadherin have been recently found \cite{Wu2015,Indra2018} (see Fig.~\ref{fig:cadherin_sketch}), and these disappear if the \textit{cis}-abolishing V81D/V175D mutation is introduced \cite{Harrison2011}. This means that while cadherin \textit{trans}-interaction mediated by membrane fluctuations does lead to the development of the larger adhesive area, it is  insufficient to help develop individual punctate cadherin adhesions.

Cadherin \textit{trans}-bond strength has been suggested to be in the range $9$-$13 k_B T$, substantially greater than the $J\le7 k_B T$ suggested for \textit{cis}-bond \cite{Wu2015}. The \textit{on}-rates of \textit{trans} bonds should therefore be $10$-$100$ times greater than those of \textit{cis} bonds, and we expect, as in much of the literature, for \textit{trans} bonds to form before \textit{cis} bonds. The problem then no longer needs to be resolved separately in both cells; rather, we need only look at the \textit{cis} (in plane) clustering of the cadherin \textit{trans}-dimers on the adhesion surface. {Recent computational work by Yu et al. \cite{Yu2022} (see their Figure 3.B) gives yet more credence to the idea that at both trans and cis-bonds are required to form punctate adherens junctions with substantial numbers of cadherins of the kind seen by Wu et al. \cite{Wu2010}, and that there is an optimum cis-bond strength for cluster formation of $3$-$8 k_B T$. We take $5 k_B T$ as a compromise value, given previous research suggesting that \textit{cis} bond strength is smaller than \textit{trans} \cite{Wu2010}.}

The effective binding strength of the each cadherin \textit{trans}-dimer unit to its neighbours is then twice as strong as the monomer value: $2 J \approx 10 k_B T$. The remainder of this work will assume that all cadherins that will form clusters have \textit{trans}-dimerised, and that the other cadherins do not impede the pair interaction.

To simplify terminology and to make the comparison with the aggregation of integrin complexes more apparent, we will henceforth use the term ``adhesion complex'' to denote a transmembrane cadherin \textit{trans}-dimer, bound to catenin and to the actin cytoskeleton insoide each cell (Fig.~\ref{fig:cadherin_sketch}).
We seek to examine the initial nucleation of cadherin clusters  revealed in super-resolution microscopy \cite{Wu2015}. This occurs before the aggregation coarsens into large adherens junctions  \cite{TruongQuang2013}. 

\section*{The model and methods}
We recently showed \cite{Ibata2020} how attractively coupled integrin adhesion complexes modelled on a 1D lattice undergo a density-dependent phase transition if their initially high concentration decreases past a critical value as the lattice length increases while their number is held constant. This was to model the development of nascent focal adhesions during cell spreading and correctly predicted the number of adhesion clusters.

The analogy between integrin and cadherin bonding suggests that a lattice gas model might be applicable to the nucleation of cadherin clusters. However, the cell-cell interface is fundamentally 2D (as opposed to integrins concentrated on the rim of the contact area), and we will find that the transition here happens as density increases, more in line with a gas-liquid condensation (again, different from the integrin transition). Accordingly, we model the contact plane between two cells as a 2D lattice with Neumann boundary conditions, with a total number of lattice sites $A$, and total area $\Sigma=a^2A$, with $a$ the lattice spacing or the distance between cadherin \textit{trans} dimers. Changes in the density of cadherins are introduced by the adiabatic change the area of the lattice.

The Hamiltonian of the attractively coupled cadherin \textit{trans}-bonded pairs is an adaptation of the Ising model:
\begin{equation}
H = - 2 J \Sigma_{\langle ij \rangle} \eta_i \eta_j \ , \label{eqn:Hamiltonian}
\end{equation}
where the sum is over nearest-neighbor pairs. The variable $\eta$ keeps track of the occupation number of each site:
\begin{align}
\eta_i = 
\begin{cases} 
     & 1 \ \ \ \ {\text{adhesion complex present} } \\
     & 0 \ \ \ \ {\text{adhesion complex absent}}
\end{cases} \ \ \ \ .
\end{align}
The binding energy $2J$ between two \textit{trans}-dimerised cadherin adhesion complexes is double the monomer binding energy $J$. We show in the Supplementary Materials that a cytoskeletal pulling force, contributiong to the Hamiltonian with a linear term $- \Sigma_i h \eta_i$, does not effect the size or distance between clusters, or the concentration at which aggregation begins.

In order to examine the spatial clustering of cadherins, we need to derive the Ginzburg-Landau action of the density distribution of cadherin units. In addition, the diffusion time over the area of membrane that separates the punctate adhesions (50-100 nm) is ca.$1$ s, similar to the growth time of the cluster (see below). The onset of aggregation is faster still, so the number of cadherins within the area which collapses into a single adhesion (the local $N/A$ value) should not change substantially at this crucial point. 


Over a sufficiently short time interval, the contact area $A$ can be assumed constant, and we need not worry about a changing expression for the partition function. The single-molecule partition function for the lattice gas model with Hamiltonian  \eqref{eqn:Hamiltonian}  is:
\begin{equation}
Z_i = \Sigma_{\eta_i = \{0,1\}}  e^{-\beta [- 2J \Sigma_{\langle j \rangle} \eta_j ] \cdot \eta_i}  \ ,
\end{equation}
where the sum runs over all of the $A$ sites in the contact area. 
The full partition function is the product of all $Z_i$, subject to the constraint of the constant total number of individual cadherin units, $N = \Sigma_i \eta_i$. The Supplementary Part B gives the calculation of this partition function using the auxilliary fields method. 
There, we introduce a site-specific variable $\rho_i$, whose expectation value is the average occupation of a site $\langle \eta_i \rangle $, and later transform this into a continuous density $\rho (s)$ which depends on a the position $s$ in the contact area.

The calculation of $Z_{\text{tot}} = \delta(\Sigma_i \eta_i - N) \Pi_{i=1}^{A} Z_i $ is exact, but rather unwieldy. We need to make two strong assumptions if we want a manageable form for the effective action $S[\rho]$. First (discussed in Supplementary Part B), the concentration of sensors is assumed low, so that the probability of a single site being in the `empty' state is much greater than that of being in the `filled' state. Second (Supplementary Part C), we look at the nucleation of clusters, where non-uniformity amplitude is small, so we can work with the series expansion of $S$ in terms of density fluctuations $\phi= \rho -N/A$.

Note that first order terms in the new variable $\phi$ average to zero, and only result in a constant shift in the action. This is why the strength of the cytoskeletal pulling force encapsulated in the field term $h$ does not change the kinetics of cadherin punctate adhesion aggregation.

In Supplementary Part D, we obtain the action $S[\rho_i]$, transform it into Fourier space, make it continuous, and finally transform back into real space (the last operation generating the spatial gradients). It has recognizable features of the Ginzburg-Landau theory, where we retain cubic and quartic terms in the order parameter expansion:
\begin{align} \label{eqn:expansionmid}
S_{\Lambda_0}[\mathbf{\phi}] = \int  d \mathbf{s} \Big[\frac{r_0}{2} \phi^2(\mathbf{s}) + \frac{c_0}{2} [\nabla \phi(\mathbf{s})]^2 + \frac{t_1}{3!}\phi^3(\mathbf{s}) 
+ \frac{u_1}{4!}\phi^4(\mathbf{s}) + \frac{u_2}{4!} \phi^2(\mathbf{s}) \Big( \int d \mathbf{s}' \phi^2(\mathbf{s}') \Big) \Big] \ ,
\end{align}
where all lengths are scaled by the size of the individual sensor $a$, and the coefficients are listed in Supplementary Part D. Specifically, the two quadratic-order coefficients take the form, Eqn.~(D.9):
\begin{equation}
r_0 = 8 \beta J  (1 - 8 g_2 \beta J)  \ ; \ \ \  c_0 =  32 g_2 \beta^2 J^2 - 2 \beta J \ ,  \label{eq:r0}
\end{equation}
with $g_2= {N}/{A}-  {N^2}/{(A+N)^2} $, from Eqn.~(C.5).

The control parameter (replacing the temperature in the classical theory of phase transitions) is the ratio $N/A$, which starts near 0 for an initially large contact plane, and then increases as the cell contact area $A$ contracts. The gradient coefficient $c_0$ remains positive, but the main `control' coefficient $r_0$ could become negative at a critical value of $N/A$ (where $g_2=k_BT/8J$) and cause the cadherin distribution to become unstable. Note that near this transition point $c_0$ takes the value $c_0 = 2 \beta J$.

\section*{Results and comparison with experimental data}


In our model, the gas-liquid condensation transition occurs when $r_0(J, N/ A) = 0$. Given $J \approx 5 k_B T$ this gives {$N/A \approx 0.05$} in a 2D lattice model of cadherin aggregation. This is our first key result, and it matches well with the \textit{in-vitro} observation that cadherin clusters form when their surface density increases past $1100 \ \text{cadherins} \ \mu \text{m}^{-2}$ \ \cite{Thompson2019} (this is different from the fraction ca.$0.01$ of the maximum cadherin surface density which they report, because they consider a much tighter packing of cadherins to within $3$ nm of their neighbours).

\vspace{0.15cm}
 \textbf{Spatial frequency of fluctuations.} \ Near the transition point, the Ginzburg-Landau action \eqref{eqn:expansionmid} can be approximated by its quadratic terms.
The time-dependence of the concentration fluctuation near the transition point can be described by the Cahn-Hilliard equation \cite{Cahn1961}:
\begin{equation}
\frac{\partial \phi}{\partial t} = D \nabla^2 \Big(r_0 \phi - c_0 \nabla^2 \phi \Big) \ \ ,
\end{equation}
where $D$ is the (not yet dimensional) cadherin diffusion coefficient in the plane. We impose Neumann boundary conditions on a 2D rectangular cell interface (a reasonable approximation for the lateral surfaces of epithelial cells for instance), which makes the time- and the spatial coordinates fully separable, and gives the Cahn-Hilliard solution:
\begin{align}
\phi(x,y,t) = \frac{1}{2}A_0(t) \label{eqn:CHsolution} 
+ \Sigma_{\mathbf{k}} A_{\mathbf{k}} (0) e^{-D \frac{\mathbf{k}^2}{4} \Big(r_0+\frac{c_0 \mathbf{k}^2}{4} \Big) t} \cos{\Big(\frac{k_x x}{2} \Big)}\cos{\Big(\frac{k_y y}{2} \Big)} , 
\end{align}
where $\mathbf{k} = (k_x, k_y)$ is the wavevector, directly related to the numbers $(m_x,m_y)$ of peaks along the two spatial directions in our contact plane: $\mathbf{k} = (2 \pi m_x/L_x,2 \pi m_y/L_y)$, where the sizes of the lattice in the x and y directions are $L_x$ and $L_y$ respectively, with $L_x L_y = A$.

The size of the fastest growing wavevector, which corresponds to the oscillation length scale that maximizes the exponential term above, are
\begin{equation}
|\mathbf{k} |_{\text{max}} =  \sqrt{{-2 r_0}/{c_0 }}\ . \label{eq:kmax}
\end{equation}

A range of mode numbers $(m_x,m_y)$ satisfy the conditions above. Assuming, as experiments suggest in Fig. 1, that the punctate adherens junctions are roughly equidistant from each other in both directions, we find that the wavenumber of the fastest-growing mode satisfies:
\begin{align}
&k_{x,\text{max}} \approx k_{y,\text{max}} \Rightarrow \frac{m_{x,y}}{L_{x,y}} = \frac{1}{2 \pi} \sqrt{-\frac{r_0}{c_0}}  .
\end{align}
The total number of clusters within the patch of cell-cell contact area  is the product $M = m_x m_y$. 

\vspace{0.15cm}
 \textbf{Mode destabilisation time.} \ A mode cannot grow until it becomes unstable, that is, when the second-order terms in the Ginzburg-Landau action become negative for that value of $k$. Lower $k$ wavevectors become unstable closer to the  density at the transition point ($r_0$), while at higher $k$ wavevectors the increasing interface energy requires a larger negative $r_0$ to destabilize the homogeneous density. Alternatively, large diffuse clusters form before smaller clusters. Assuming a steady rate change of the contact area, we find in Supplementary Part F that the largest number of clusters $M$ in the lattice, for which the combined second order Ginzburg-Landau term is negative, is proportional to the time $t_1$ that has elapsed since the cadherin density crossed the transition point $r_0 = 0$. Reintroducing dimensional length scales, we find this linear relation:
\begin{align}
M \approx \Psi(\beta J) \frac{t_1 | \dot{\Sigma}|}{a^2} \ ,  \label{K1}
\end{align}
where $\dot{\Sigma}$ is the rate of decrease in the cell-cell contact area, and $\Psi(\beta J)$ is a complicated function obtained from the series expansion, Eqn. (F.2).

\vspace{0.15cm}
 \textbf{Mode growth time.} For the fastest growing mode, $r_0 = -c_0 k_{\text{max}}^2/2$, Eqn. \eqref{eq:kmax}, so we find  that the mode has a characteristic exponential growth time $t_2$ which dictates how fast the Ginzburg-Landau free energy is minimized. We find this time by substituting into the exponent in the Cahn-Hilliard solution  \eqref{eqn:CHsolution}:
\begin{equation}
e^{-D \frac{k^2}{4} \Big(r_0 + \frac{c_0 k^2}{4} \Big) t}\Big|_{\text{max}} = e^{D \frac{k^4 c_0}{16} t}
\end{equation}
Substituting the constants $r_0$ and $c_0$, and recovering the proper dimensional length scales via the spacing $a$, we find that the growth time of the mode becomes:
\begin{equation}
t_2=\alpha \frac{A^2 a^2}{2^6 \pi^4 M^2 D  g_2 \beta^2 J^2} \ , 
\label{K2}
\end{equation}
i.e. inversely proportional to $M^2$, with $\alpha$  a proportionality constant of order $1$. 

\vspace{0.15cm}
\textbf{Density of punctate adherens junctions.} \ The total time for a mode to first become unstable and then reach the minimum in the Ginzburg-Landau free energy landscape is of the order:
\begin{equation}
t_{\text{tot}} = t_1 + t_2 = K_1 M + K_2 M^{-2}  \label{eq:2times}
\end{equation}
where $K_1$ and $K_2$ are the proportionality constants in Eqs.~(\ref{K1},\ref{K2}). The total time is minimized when the number of clusters across the contact area is:
\begin{align}
\frac{d t_{\text{tot}}}{d M} &= K_1 - 2 K_2 M^{-3} = 0  
\Rightarrow M^* = \Big(\frac{2 K_2}{K_1} \Big)^{1/3}  .
\end{align}
 We make the proportionality constants explicit and find that the number  of adherens junctions per unit area of cell-cell interface is given by
\begin{equation}
n_{\text{s}} = \frac{M^*}{a^2 A} = \Big( \alpha \frac{\Psi(\beta J)}{2^5 \pi^4 a^4 D g_2 \beta^2 J^2} \frac{|\dot{\Sigma}|}{\Sigma}\Big)^{1/3} \label{eqn:propm}
\end{equation}
This is the second main prediction of this paper.

\begin{table*}[t]
\begin{center}
\begin{tabular}{ c | c | c | c | c }
 Parameter & Name & Value & Uncertainty & References \\ 
 \hline
Cadherin lattice \\ spacing & a & 7 nm & $\pm$ 1 nm & \cite{Patel2006,Wu2010,Harrison2011,Thompson2019} \\  
Cadherin diffusion \\ coefficient (macroscopic) & D & $ 2.6\cdot 10^{-3} \mu \text{m}^2/$s & $\pm 1.1\cdot 10^{-3} \mu \text{m}^2/$s & \cite{Kusumi1993}  \\
Cadherin diffusion \\ coefficient (microscopic) & D & $ 3.3\cdot 10^{-3} \mu \text{m}^2/$s & $\pm 2.9\cdot 10^{-3} \mu \text{m}^2/$s & \cite{Kusumi1993}  \\
Cis interaction energy & J & $ 5 k_B T$ & $\pm 2 k_B T$ & \cite{Wu2010, Yu2022} \\
Fractional shrinkage of \\ cell-cell contact area & $|\dot{\Sigma}|/\Sigma$ &  $\approx 20 \% 
\text{min}^{-1}$ & $\pm 5 \% \text{min}^{-1}$ & Figure 2 \cite{Guillot2013} \\
Density at the phase transition & N/A & 0.05 & $ \pm 0.01 $ & (see above and \cite{Thompson2019}) \\
Arbitrary exponential growth parameter & $\alpha$ & 1-10 & $ * $ & NB$\dagger$
\end{tabular}
\end{center}
\caption{{Summary of the physiological values used in the model. Note the large standard deviation in the cadherin diffusion constants. This arises due to the different modes of cadherin movement \cite{Kusumi1993}; however, the uncertainty on the average value is much less than this standard deviation would suggest. $\dagger$ NB: Depends on the tolerance on how far the coarsening has progressed. Here we choose a factor of $e$ (ca. $63\%$ growth), whereas simulations with a $1\%$ tolerance \cite{Konig2021} give a value of $\alpha$ closer to 10. It seems clear that cadherin clustering should become visible before coarsening completes.}}
\end{table*}

To make an experimental comparison, we use the aggregated cadherin spacing $a$ = 6 nm \cite{Patel2006,Wu2010,Harrison2011} (slightly smaller than the crystal lattice size seen by \cite{Harrison2011} to account for the possible tighter packing observed by \cite{Thompson2019}), diffusion constant $D = 3\cdot 10^{-3} \mu \text{m}^2/$s \cite{Kusumi1993}, density at transition of $N/A = 0.04$ (see above), $J=5 k_B T$, and proportionality constant $\alpha = 1$ for the sake of simplicity (see Table 1). The fractional shrinking rate of a patch of the cell-cell contact area $(|\dot{\Sigma}|/\Sigma)$ depends on the cellular process and the type of cell.

\begin{figure}
 \centering
\includegraphics[width=0.5\textwidth]{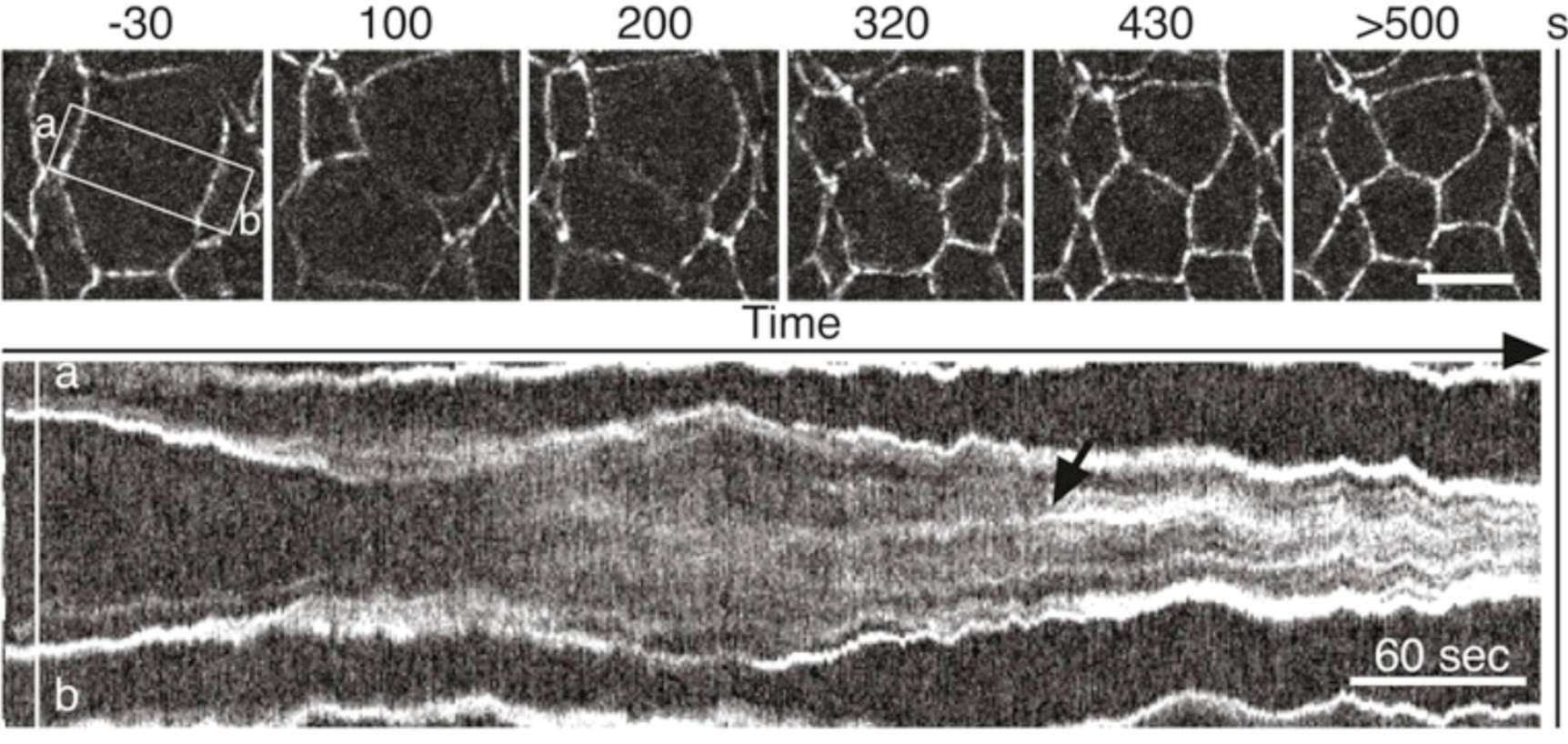}
\caption{Evolution of cadherin density on the apical edge of a dividing epithelial cell during cytokinesis, reproduced with permission from \cite{Guillot2013}. Cadherins are labelled with E-cad::GFP fluorescent markers; the second panel (kymograph) shows the density along the [a-b] cross-section, illustrating how the interface area contracts and gets populated by increasingly inhomogeneous cadherin clusters.  Aggregation begins only after $200$s; during this time, the length of the apical edge of the new contact area between daughter cells reduces at a rate of $|\dot{\Sigma}|/\Sigma \approx 20 \% \ \text{min}^{-1}$.}\label{fig:Fig2_cadh}
\end{figure}

One classic example which requires cadherin clusters to form occurs during cytokinesis at the end of cell division \cite{Barr2007}. Kymographs of the cross-section of the a dividing epithelial cell \cite{Guillot2013,Herszterg2013}, see Fig.~\ref{fig:Fig2_cadh}, show that the fractional change in the area of the cell interface is $|\dot{\Sigma}|/\Sigma \approx 20 \%$ per minute. For the case of epithelial cell division, therefore, we find {$n_{\text{s}} \approx 28 \pm 4 \ \text{clusters} \ \mu \text{m}^{-2}$, assuming a $40 \%$ uncertainty for the cadherin diffusion coefficients (macroscopic uncertainty value in Table 1).}. This appears to match well with the number density of punctate adhesions observed by \cite{Wu2015} in epithelial tissues seen in Fig. \ref{fig:cadherin_sketch}. 
Both experimental and computational studies have also reported the characteristic cluster size values of $33$ cadherins \cite{Thompson2020}, which gives the estimate of average cadherin of the correct order of ca.$1000 \ \text{cadherins} \ \mu \text{m}^{-2}$ \ \cite{Thompson2019}.
Even though there might be some uncertainty in the values of physiological constants used to evaluate Eqn.~\eqref{eqn:propm}, the cube-root dependence of the number density on these constants makes a large error unlikely: any one parameter value would need to be off by a factor of $1000$ for there to be $10$ times fewer or more punctate adhesions. {Note that we can also use this figure to estimate the total growth and destabilisation time to be of the order of 5 seconds, once the cadherin distribution becomes unstable. This would account for the quite sharp transition to a higher cadherin instensity in the kymograph at the time indicated by the arrow.}

\section*{Conclusions}
In this paper, we built a 2D model for the aggregation of adhesion units with a short-distance attractive interaction, which is applicable to all cell contact areas if the problem is reduced down to a sufficiently small and uniform patch (locally, with zero curvature and applied force variation), so we expect our results to be  more universally valid. The advantage of treating cadherin adhesion complexes as a specific example is that there is a large body of experimental and computational work from which to test our results. Our predictions for the transition density of cadherins above which clusters could form, as well as for the number density of cadherin clusters, were independent of each other. Together, they strongly suggest that cadherin density-depedent cluster nucleation initially occurs via a gas-liquid phase transition.

{Previous studies have analytically explained how cadherin trans bonds can help set up large scale adherens junctions\cite{Fenz2017}, and new work has computationally shown that cadherin cis-bonds help develop punctate adherens junctions\cite{Yu2022}. In this work. we have laid out for the first time an \textit{analytical} explanation for how \textit{punctate} cadherin adherens junctions can form.}

While we looked at the case of cadherin nucleation, more generally, any distribution of attractively-coupled cell membrane molecules in the low concentration limit can undergo a gas-liquid (condensation) phase transition in the form of density-dependent aggregation. Because of this, we could apply this method to the more general problem of the formation of membrane-bound organelles. 

\subsection*{Data access}
All data generated or analysed during this study are included in this published article [and its supplementary information files].

\subsection*{Author contribution}
Both authors conceived the idea, carried out different elements of data analysis and wrote the paper.

\subsection*{Acknowledgements}
We are grateful to R. Zaidel-Bar as well as to T. Lecuit for giving access to their experimental results. This work has been funded by BBRSC DTP Cambridge (grant no. EP/M508007/1).


\end{document}